\documentclass[aps,prd]{revtex4}
\usepackage{graphics,psfrag}
\usepackage{graphicx,psfrag}
\usepackage{epic,eepic}
\usepackage{amsmath}
\usepackage{stmaryrd}
\usepackage{color} 
\usepackage{epsfig}
\usepackage{latexsym}
\usepackage{pstricks}
\newcommand{\be}{\begin{equation}}
\newcommand{\ee}{\end{equation}}
\newcommand{\bear}{\begin{eqnarray}}
\newcommand{\eear}{\end{eqnarray}}
\begin{document}

\title{Vacuum quark condensate, chiral Lagrangian, and Bose-Einstein statistics}
\author{Taekoon Lee}
\email{tlee@kunsan.ac.kr}
\affiliation{Department of Physics and Astronomy, 
Johns Hopkins University, Baltimore, MD 21218, USA and \\
Department of Physics, Kunsan National University, 
Kunsan 573-701, Korea\footnote{\!\!Permanent address.}}


\begin{abstract}
In a series of articles it was recently claimed that the quantum chromodynamic (QCD)
 condensates are
not the properties of the vacuum but of the hadrons and are confined inside them.
 We point out that this claim is incompatible with the chiral Lagrangian and 
 Bose-Einstein statistics 
of the Goldstone bosons (pions) in chiral limit
 and conclude that the quark condensate must be the property of the QCD vacuum. 

\end{abstract}

 

\maketitle

Spontaneous symmetry breaking (SSB) is the fundamental concept that pervades
many disciplines of physics, from ferromagnetism  
and superconductivity in solid state physics to the chiral symmetry braking
in QCD to the Higgs mechanism in the standard model of electroweak interactions.
In conventional wisdom of SSB the vacuum (ground state) plays the most
crucial role; SSB occurs when the vacuum does not respect the 
 symmetry of the theory by developing an order parameter  that is
 not invariant under the symmetry.  

In a series of articles it was recently argued  that this conventional wisdom does
 not apply to the chiral symmetry breaking of QCD 
 \cite{brodsky08-0,brodsky08-1,brodskyplb,brodsky09-1,brodskyPRC0,brodskyPRC1}.
  Invoking the confinement of quarks and
 gluons as well as the supposed triviality of the Fock vacuum  in the light-front formalism of QCD
 it was argued that the chiral symmetry breaking  is the property of hadron dynamics and
 occurs only inside the hadrons.  The consequence of this is that the quark condensate is 
 confined inside the hadrons, leaving
 the QCD vacuum {\it empty}, devoid of vacuum condensates that fill the space-time. 
  With no order parameter in the vacuum, 
  even in broken phase the vacuum would be invariant
  under the  chiral symmetry, which is in a stark contrast to the conventional wisdom. 
  This view was recently criticized in \cite{reinhardt}, which was 
  counterargued in \cite{brodsky12-0}.
  
  If the claim were true, it would not only demand a fundamental change in
  our understanding of SSB but also have important phenomenological implications;
  The absence of the vacuum condensates would 
   eliminate the apparent enormous contribution of the vacuum condensates
   to the cosmological constant, and this was suggested as a 
   solution to the cosmological constant problem \cite{brodsky08-0,brodsky08-1,brodsky09-1}.
   But the gravitational effect is
   not the only observable channel because the absence of the vacuum condensate can also be 
  directly probed by the 
  disoriented chiral condensate effect in heavy-ion collisions
   \cite{bjorken,blaizot,nelson,anselm,dcc}, which occurs when a
  quark-gluon plasma cools down and 
  develops quark condensate that is misaligned from
  the ambient vacuum quark condensate.
   The eventual aligning to the 
  vacuum condensate of the misaligned condensate causes 
  coherent production of pions. However, with the absence 
  of vacuum quark condensate there would be no such effect.
 
  In this note we  point out that  
  the notion of in-hadron condensate is incompatible with 
  the well-established chiral Lagrangian as well as the
   Bose-Einstein statistics
  of the pions in chiral limit, thus invalidating  the claim.
  

  The argument for in-hadron condensates is largely based on the observation that the
  vacuum quark condensate in the Gell-Mann--Oakes--Renner relation can be
  expressed in terms of the pion-to-vacuum transition amplitude,
  \bear
  \langle\bar q q\rangle\sim f_{\pi}\langle 0|\bar q\gamma_5 q|\pi\rangle\,,
  \label{con}
  \eear
  where $f_\pi$ is the pion decay constant,
  which combined with the confinement of quarks and gluons was used to interpret
  the vacuum quark condensate as the property of the pion 
  wave function in Bethe-Salpeter equation
  rather than that of the vacuum. 
  However, (\ref{con})
  is  the consequence of chiral symmetry breaking and does 
  not preclude nonzero vacuum quark condensate.
   In fact, in any SSB the order parameter 
   can be expressed in terms of the Goldstone boson-to-vacuum 
   transition amplitude via insertion of the operator that
   defines the order parameter, which comes from the 
   fact that the Goldstone bosons 
   are zero-mode excitations
   about the vacuum condensate.
  
  It is instructive to derive (\ref{con}) using the 
  chiral Lagrangian for pions
  \bear 
  {\cal L}= \frac{f_{\pi}^2}{4} {\rm Tr} \partial\mu 
  \Sigma^\dagger \partial^\mu \Sigma + \alpha ({\rm Tr}M\Sigma+h.c.)
  \eear
  where $M$ denotes the current quark mass matrix and $\alpha$ 
  is a dim-3 constant that is, by definition, positive.
   Under a two-flavor chiral symmetry
  $SU_L(2)\times SU_R(2)$ the chiral field $\Sigma$ 
  transforms as $\Sigma \to V\Sigma U^\dagger$
  where $U \in SU_L(2), V\in SU_R(2)$.
  
  Putting 
  \bear
  \Sigma=\Sigma_0 e^{i2\Pi/f_\pi}
  \eear
  where $\Sigma_0 \in SU(2)$ and $\Pi=\pi^a \sigma^a/2$, 
  with $\sigma^a$ denoting the Pauli matrices,
  we obtain
  \bear
  {\cal L} =\alpha ({\rm Tr} M\Sigma_0+h.c) +\frac{1}{2}
  \partial_\mu \pi^a \partial^\mu \pi^a  -\frac{\alpha}{2f_\pi^2}(\pi^a)^2
  ({\rm Tr}M\Sigma_0 +h.c.) +O(\Pi^4)\,,
  \label{chiral1}
  \eear
  where  the 
  reason for the missing  linear term in pion fields shall be given shortly.
  Now the vacuum alignment by Dashen \cite{dashen}
   suggests   $\Sigma_0$ be such that it minimizes the 
   quark-mass induced Hamiltonian term
  \bear
  H_M=-\alpha ({\rm Tr} M\Sigma+h.c.)\,.
  \label{massterm}
  \eear
   $\Sigma_0$ is thus the chiral field  configuration that
  minimizes the vacuum energy of the chiral theory.  
  Moreover, since the vacuum quark condensate of QCD in chiral limit can be defined as
  \bear
  \langle \bar q_{L i} q_{R j}\rangle= \left.-i\frac{\partial 
  \ln {\cal Z}_{QCD}(M)}{V\partial M_{ij}}\right|_{M=0, V\to\infty}\,, 
  \label{condensate0}
  \eear
  where ${\cal Z}_{QCD}$ denotes the QCD partition function, $V$  the space-time volume,
   and invoking the quark-hadron duality to substitute the QCD
  partition function with that of the chiral Lagrangian we get from (\ref{chiral1})
  \bear
  \langle \bar q_{Li} q_{Rj}\rangle= \alpha \Sigma_{0ji}\,,
  \eear
 which shows $\alpha\Sigma_0$ is  nothing but the vacuum quark condensate in chiral limit.
  Since the most general $M$ can be written  as
  \bear
  M=U_0 M_D V_0^\dagger
  \label{mass}
  \eear
  where $M_D={\rm Diag}(m_u,m_d)$ with $m_{u,d}>0$, and $U_0, V_0 \in SU(2)$, 
  the chiral configuration that minimizes (\ref{massterm})
  is given by 
  \bear
  \Sigma_0=V_0 U_0^\dagger\,,
  \label{condensate1}
  \eear
   which, upon substituting into (\ref{chiral1}), yields the Gell-Mann--Oakes--Renner relation
  \bear
  m_\pi^2&=&\frac{\alpha}{f_\pi^2}({\rm Tr} M\Sigma_0+h.c.)=\frac{2\alpha}{f_\pi^2}(m_u+m_d)\,,
   \eear
   where
   \bear
   \alpha &=& \sqrt{{\rm Tr} (\langle \bar q_L q_R\rangle \langle \bar q_R q_L\rangle)/2} \,. 
   \eear
  Clearly, the condensate $\Sigma_0$ depends on the chiral direction
  of the quark-mass $M$ before it being taken to zero. Note also that  (\ref{mass}),(\ref{condensate1}) give vanishing
  linear term in $\pi^a$ fields in
  expansion (\ref{chiral1}).
   This selection of the chiral vacuum configuration
    out of the continuum of vacua by minimization of the quark-mass induced
  vacuum energy is the chiral Lagrangian version of the Dashen's vacuum alignment \cite{lee,creutz}. 
Now note that the discussion  so far does show that 
quark-hadron duality gives a nonvanishing vacuum quark condensate  
in chirally broken phase, which contradicts the claim of vanishing vacuum condensate.
The chiral Lagrangian, tested extensively, is based 
on the general principle of symmetry, and  is widely believed to be
 a valid effective theory for the pions, but the notion of in-hadron condensate is 
incompatible with it because the chiral
Lagrangian  is build on a nonvanishing vacuum quark condensate.

With the chiral Lagrangian  the quark condensate can also be easily expressed in 
terms of the pion-to-vacuum transition amplitude by
\bear
\langle 0|\bar q_L \sigma^a q_R|\pi^a\rangle= \alpha \langle 0|{\rm Tr} \sigma^a
 \Sigma |\pi^a\rangle=i\frac{3\alpha}{f_\pi}{\rm Tr} \Sigma_0=i\frac{3}{f_\pi}\langle \bar q_L q_R\rangle\,,
\label{transition}
\eear
which clearly shows that (\ref{con}) is a consequence of
 the chiral symmetry breaking and in itself does not 
imply vanishing of vacuum quark condensate. Thus brought in by the authors 
\cite{brodsky08-0,brodsky08-1,brodskyplb,brodsky09-1,brodskyPRC0,brodskyPRC1} 
is the confinement of QCD which was argued to limit the range of
quark and gluon fluctuations to regions inside the hadron, rendering the 
hadrons be a system analogous to an isolated finite volume
ferromagnet or superconductor. As such systems have condensate inside the 
finite volume it was argued that QCD 
condensate, defined through the pion-to-vacuum transition
 amplitude (\ref{con}), is confined similarly inside the hadron.
In this analogy the valence quarks play the role of the 
atoms in the condensed matter systems, and as there would be no SSB without the atoms
chiral symmetry breaking would be impossible without valence quarks,
 hence no chiral symmetry breaking in the vacuum.

  The problem of this view that each hadrons are isolated systems in an {\it empty} vacuum 
  is that the quark condensates  of each pions in chirally symmetric QCD ($M\!=\!0$)
  can have random directions in chiral space, as there is no unique limit 
  for the in-hadron quark condensate in chiral limit. 
  The possibility of the pions having differing in-hadron condensates
   and its potential 
  effect similar to the disoriented chiral condensate
   were already pointed out in \cite{brodsky08-0,brodsky09-1}. 
   However, if the pions indeed had random 
   condensates then they would be no longer identical
   particles and thus violate the Bose-Einstein statistics. 
   Therefore, all pions must have the same quark condensate, 
   but then without vacuum quark condensate
  it would have to be imposed in an ad hoc manner.
  Which is a critical flaw because the statistics 
  of particles should arise
  naturally from the theory, as in the chiral Lagrangian 
  where it is the consequence of second quantization, rather than
  arbitrarily forced. Thus, 
  the Bose-Einstein statistics also refutes the notion of in-hadron condensate.

So far, we focused on the quark condensate but it was 
also suggested that QCD confinement
 makes all vacuum condensates vanish as well 
 \cite{brodsky08-0,brodsky08-1,brodskyplb,brodsky09-1,brodskyPRC0,brodskyPRC1}.
A notable example is the gluon
 condensate $\langle G_{\mu\nu}^2\rangle$ which appears in QCD
 sum rule via operator production expansion. The scattered values of the condensate from
  various fittings, which vary from negative to positive, was 
 cited as a support for the vanishing gluon condensate \cite{brodsky08-0}. However, 
 the wide variations of the fitted values may arise from inadequate treatment of
 the perturbative contribution, which, being an asymptotic 
 series, has power correction-like contribution
 when resummed. The usual sum rule practice of truncating 
 the perturbative series at first few terms is
 based on the assumption that the nonperturbative 
 power correction is overly dominant over the 
 perturbative one \cite{david1,beneke}, but the scattered values
 may rather be an indication of the failure of 
 this assumption than  the vanishing of the vacuum condensate.
  On the other hand, direct extraction from the lattice simulation of the plaquette 
 indicates a nonvanishing gluon condensate \cite{lee2,rakow}, and there is also a theoretical 
 reason that supports nonvanishing gluon condensate: the renormalon. 
  It is well known that the power correction, 
 for instance in the Adler function, by
  the condensate is ambiguous, which ambiguity is cancelled by
the corresponding ambiguity in the Borel summation of the  perturbative series \cite{david,david1}. 
Thus, a divergent perturbative series by the renormalon would require a nonvanishing 
vacuum condensate. Indeed, the calculations for the
 normalization constant of the large order behavior of the Adler function  give 
  nonvanishing values \cite{lee1,lee3,jamin}, which again  implies that the vacuum 
 gluon condensate should not vanish. 

To conclude, we have provided following reasons to refute 
the notion of in-hadron condensates: (i) Dashen's vacuum alignment
implemented on the chiral Lagrangian gives nonvanishing vacuum quark 
condensate, which renders the notion of in-hadron condensate
incompatible with the well-established chiral Lagrangian, 
and (ii) the pions having random in-hadron quark 
condensates in chiral limit is incompatible with the Bose-Einstein statistics.

 \begin{acknowledgments}
The author is deeply grateful for hospitality to the theory group, 
and especially to Kirill Melnikov, in the 
department of physics and astronomy at Johns Hopkins University.
\end{acknowledgments}

\bibliographystyle{apsrev}  
\bibliography{qcdCondensate}

\end{document}